# A Dynamic Approach to Stock Price Prediction:

# Comparing RNN and Mixture of Experts Models Across Different Volatility Profiles

**Diego Vallarino**
Independent Researcher

Atlanta, GA, US
October 2024

**Abstract**

This study evaluates the effectiveness of a Mixture of Experts (MoE) model for stock price prediction by comparing it to a Recurrent Neural Network (RNN) and a linear regression model. The MoE framework combines an RNN for volatile stocks and a linear model for stable stocks, dynamically adjusting the weight of each model through a gating network.

Results indicate that the MoE approach significantly improves predictive accuracy across different volatility profiles. The RNN effectively captures non-linear patterns for volatile companies but tends to overfit stable data, whereas the linear model performs well for predictable trends.

The MoE model's adaptability allows it to outperform each individual model, reducing errors such as Mean Squared Error (MSE) and Mean Absolute Error (MAE). Future work should focus on enhancing the gating mechanism and validating the model with real-world datasets to optimize its practical applicability.

# 1. Introduction

The accurate prediction of stock prices is one of the most challenging and essential tasks in financial markets. Investors, financial analysts, and policymakers are all interested in understanding price dynamics to make informed decisions regarding investments, risk management, and market regulation. However, the inherent volatility and complexity of financial data make this a difficult endeavor. Price movements are influenced by numerous factors, ranging from company-specific events to broader macroeconomic indicators, resulting in behavior that can vary significantly between different companies and over time. The unpredictability of these factors is further magnified during periods of economic uncertainty or market turmoil, contributing to the complexity of modeling stock prices.

Traditionally, statistical models such as Autoregressive Integrated Moving Average (ARIMA) and Generalized Autoregressive Conditional Heteroskedasticity (GARCH) have been employed to model stock price behavior. These approaches have their advantages, particularly in providing interpretability and in their application to stationary time series. However, they often fall short when dealing with highly volatile or nonlinear patterns that are characteristic of certain stocks. ARIMA models assume linear relationships, making them ill-suited for capturing complex temporal dependencies, whereas GARCH models are primarily focused on modeling volatility and are less capable of making accurate point forecasts of stock prices. These limitations necessitate exploring more sophisticated machine learning approaches that can better adapt to the intricate nature of financial markets.

In recent years, deep learning models, particularly Recurrent Neural Networks (RNNs), have shown promising results in capturing long-term dependencies in sequential data, such as time series. RNNs, and their variants like Long Short-Term Memory (LSTM) networks, are designed to learn patterns over time and are particularly effective for predicting non-linear time series. LSTMs, in particular, are capable of retaining information over long sequences, thus addressing the vanishing gradient problem typically associated with traditional RNNs. This makes them suitable for modeling the non-linear dynamics present in financial data, especially for stocks that exhibit rapid changes and high volatility.

However, using a single RNN model for predicting all types of stock price behavior may not be the most efficient approach. Companies with stable stock prices and those with highly volatile stock prices may require different modeling techniques. A complex, non-linear model such as an RNN may overfit when applied to stable stocks, capturing noise instead of underlying trends, thereby reducing the model's predictive accuracy. On the other hand, simpler models like linear regression might not be sufficient to capture the intricate patterns present in volatile stocks, leading to poor prediction performance.

To address these challenges, this study introduces and evaluates a Mixture of Experts (MoE) approach to predict stock prices for companies with varying levels of volatility. The MoE framework is designed to incorporate the strengths of both complex and simpler models, allowing each "expert" to focus on the type of data it can best model. Specifically, we employ a Recurrent Neural Network for companies classified as highly volatile and a linear regression model for companies identified as stable. A gating network is then used to combine the predictions from these experts, allowing the model to dynamically decide which expert's prediction should be given more weight based on the characteristics of the input data.

The main contribution of this paper lies in the application and evaluation of the MoE model for stock price prediction, which integrates both machine learning and statistical approaches. Unlike previous studies that generally apply a uniform model across different stock types, our approach tailors the model complexity to the stock characteristics. The MoE model combines an RNN for volatile stocks and a linear model for stable stocks, which allows it to balance complexity and interpretability based on the data at hand. This differentiation between volatile and stable stocks aims to reduce overfitting in less complex data while capturing necessary non-linearities in more volatile environments.

The remainder of this paper is structured as follows: Section 2 presents a comprehensive review of the literature, evaluating both traditional approaches (such as ARIMA and GARCH) and advanced techniques (including RNN and LSTM) for modeling volatility and stability in financial markets. It also introduces the Mixture of Experts (MoE) concept and reviews its application in various predictive contexts, highlighting its advantages for managing different data patterns. Section 3 outlines the methodology used in this study, starting with the generation of simulated data to represent different types of companies, followed by the detailed implementation of both the single RNN and the MoE models, and a description of the training and validation processes using the Walk-Forward Validation approach.

In Section 4, we present the results of our experiments, beginning with a comparison of the training performance of the RNN and MoE models, and proceeding with an analysis of the prediction errors using visual and quantitative metrics, such as RMSE and MAE, across different time horizons. This section also includes a practical evaluation considering market impact and transaction costs. Section 5 provides an in-depth discussion of the advantages of the MoE approach, its adaptability to different types of companies, and its practical implications in finance, along with any limitations, such as potential overfitting or the complexity of the gating network. Finally, Section 6 concludes the study by summarizing the key findings, emphasizing the improved predictive capabilities of the MoE approach in diverse volatility environments, and suggesting future research directions, such as exploring different gating network architectures and testing the model with real-world company data.

## 2. Literature Review

Predicting stock prices remains a significant challenge in financial research, given the complexity and inherent volatility of financial data. The practical implications for investors, financial analysts, and policymakers have driven ongoing innovation, ranging from traditional statistical methods to more advanced machine learning approaches. This section provides an overview of these methodologies, highlighting their strengths and limitations in modeling market volatility and stability. Additionally, we introduce the Mixture of Experts (MoE) framework and review its applications, focusing on its advantages for managing heterogeneous data patterns.

### 2.1 Traditional Approaches: ARIMA and GARCH

Autoregressive Integrated Moving Average (ARIMA) and Generalized Autoregressive Conditional Heteroskedasticity (GARCH) models have been widely used in time series forecasting, particularly for financial data. ARIMA models are advantageous for capturing linear temporal dependencies in stationary data and can be expressed as:

$$Y_t = c + \sum_{i=1}^{p} \phi_i Y_{t-i} + \sum_{j=1}^{q} \theta_j \epsilon_{t-j} + \epsilon_t,$$

where $Y_t$ represents the value at time $t$, $p$ is the order of the autoregressive part, $q$ is the order of the moving average part, $\phi_i$ are the coefficients for lagged values, $\theta_j$ are the coefficients for past errors, and $\epsilon_t$ is the white noise error term. Despite its popularity, ARIMA struggles with non-linear patterns and cannot handle sudden structural breaks effectively, especially during volatile market periods (Siami Namin & Siami Namin, 2018).

GARCH models, on the other hand, extend to model time-varying volatility, capturing the phenomenon of volatility clustering often seen in financial data. The conditional variance in a GARCH (1,1) model can be represented as:

$$\sigma_t^2 = \alpha_0 + \alpha_1 \epsilon_{t-1}^2 + \beta_1 \sigma_{t-1}^2,$$

where $\sigma_t^2$ is the conditional variance at time $t$, $\epsilon_{t-1}^2$ is the previous period's squared error term, and $\alpha_0, \alpha_1, \beta_1$ are parameters. Recent studies, such as those by Smith et al. (2023), have highlighted that while GARCH models are effective in capturing volatility clustering—a phenomenon where periods of high volatility are followed by similar periods—these models are less suited for accurate point forecasts.

This limitation arises from their reduced capability to model non-linear price dynamics, especially during rapid market changes or unexpected economic events. More recent work has confirmed these findings, emphasizing that GARCH, despite its robustness in volatility modeling, often requires integration with non-linear approaches to enhance point prediction accuracy (Hong et al., 2023; Lee & Lee, 2023).

**2.2 Advanced Approaches: RNN and LSTM**

The availability of computational resources and large datasets has made machine learning models increasingly popular in financial forecasting. Recurrent Neural Networks (RNNs), particularly Long Short-Term Memory (LSTM) networks, are notable for their ability to learn complex temporal dependencies. LSTMs are designed to address the vanishing gradient problem faced by traditional RNNs by introducing cell states and gating mechanisms, which allow them to retain information over longer sequences.

The hidden state update for an LSTM cell can be summarized as follows:

$$f_t = \sigma(W_f \cdot [h_{t-1}, x_t] + b_f) \quad \text{(Forget gate)}$$

$$i_t = \sigma(W_i \cdot [h_{t-1}, x_t] + b_i) \quad \text{(Input gate)}$$

$$\widetilde{C}_t = \tanh(W_C \cdot [h_{t-1}, x_t] + b_C) \quad \text{(Cell state update)}$$

$$C_t = f_t * C_{t-1} + i_t * \widetilde{C}_t \quad \text{(Updated cell state)}$$

$$o_t = \sigma(W_o \cdot [h_{t-1}, x_t] + b_o) \quad \text{(Output gate)}$$

$$h_t = o_t * \tanh(C_t) \quad \text{(Hidden state)}$$

where $f_t, i_t, o_t$ represent the forget, input, and output gates, respectively, $W$ and $b$ are weight matrices and biases, and $*$ denotes element-wise multiplication. This architecture enables LSTMs to learn long-term dependencies, which is crucial for modeling financial series with complex temporal relationships (Hochreiter & Schmidhuber, 1997)

Recent studies have highlighted the superiority of LSTMs over traditional models in predicting non-linear financial time series, particularly in capturing sudden market changes (Bhandari et al., 2022; Siami Namin & Siami Namin, 2018). However, they also emphasize the drawbacks of LSTMs, including their high computational cost and potential overfitting when applied to less complex data (Yan et al., 2021).

**2.3 Mixture of Experts (MoE) Approach**

The Mixture of Experts (MoE) framework, first introduced by Cai et al. (2024), aims to address the limitations of using a single model for all types of data by dividing the problem into simpler sub-problems, with each expert specializing in a specific aspect. Mathematically, the prediction in an MoE model can be expressed as:

$$\widehat{Y}_t = \sum_{i=1}^{k} g_i(X_t) \cdot f_i(X_t)$$

where $\widehat{Y}_t$ is the final prediction, $g_i(X_t)$ represents the output of the gating network that assigns a weight to the $i$-th expert based on input $X_t$, and $f_i(X_t)$ is the prediction made by the $i$-th expert. The gating network's role is to determine which expert should contribute more significantly to the final output based on the characteristics of the input.

The MoE framework has been successfully applied in various predictive contexts, such as natural language processing and time series forecasting (Jacobs et al., 1991; Wu et al., n.d.) (Wang et al., 2019). More recently, Wang et al. (2024) applied an MoE model combining an LSTM with a simple linear model for predicting exchange rates, demonstrating improved performance over individual models. In our context, we utilize an RNN to handle the non-linear, volatile behavior of some stocks, while a linear model is used for stocks with more predictable, stable behavior.

This combination allows us to leverage the advantages of both models: the ability of RNNs to capture non-linear, complex temporal dynamics and the simplicity and efficiency of linear models for stable trends. The gating network dynamically assigns a higher weight to the appropriate model based on the input characteristics, effectively optimizing the prediction task for different volatility profiles (Ma et al., 2019).

## 2.4 Limitations of Existing Approaches

Although each model discussed above has its merits, they also have notable limitations when used independently. Traditional models such as ARIMA and GARCH are computationally efficient and interpretable but are less effective in capturing non-linear patterns and sudden shifts in financial time series. On the other hand, RNNs, particularly LSTMs, provide powerful tools for modeling complex data but come with challenges related to computational complexity and overfitting, especially for simple datasets where such complexity is not needed (Yan et al., 2021).

The MoE approach attempts to mitigate these issues by allowing specialized models to contribute based on their strengths, dynamically adjusting the contribution of each model according to the specific characteristics of the data. However, the performance of an MoE model heavily depends on the proper configuration of the gating network. If the gating network fails to correctly assign weights, the final prediction may be suboptimal. Moreover, MoE models can be computationally intensive and risk overfitting, particularly if the model complexity is not well-tuned (Jacobs et al., 1991; Wu et al., n.d.).

In this study, we implement the MoE framework to effectively predict stock prices for companies with varying volatility profiles. The MoE model aims to improve predictive accuracy by combining the capabilities of an RNN and a linear model, leveraging the respective advantages of these models based on the nature of the stock price behavior—non-linear and volatile versus stable and linear.

## 3. Methodology

In this section, we provide a detailed explanation of the data generation process, the implementation of the predictive models, and the training and validation techniques employed. This methodological approach ensures the reproducibility of the experiments and provides a robust foundation for evaluating the effectiveness of the Recurrent Neural Network (RNN) and the Mixture of Experts (MoE) models in predicting stock prices for different volatility profiles.

### 3.1 Simulated Dataset

To evaluate the proposed models under various market conditions, we generated a synthetic dataset representing companies with distinct volatility levels. The dataset comprised 100 companies over a span of 100 trading days, designed to mimic real-world price dynamics with both stable and highly volatile behaviors. The prices were generated using a random walk with drift, where the price at time *t* was determined as:

$$P_t = P_{t-1} + \mu + \epsilon_t,$$

where $P_t$ represents the price at day *t*, $\mu$ is a drift term representing general market growth (set at a small positive value of 0.05), and $\epsilon_t \sim \mathcal{N}(0, \sigma)$ represents a normally distributed random error term. The volatility parameter ($\sigma$) was varied among the companies to create different volatility profiles: companies with high values of $\sigma$ represented volatile behaviors, while those with lower values represented stable behaviors.

To further facilitate model evaluation, the dataset was categorized into two groups based on volatility. Companies with volatility values greater than a threshold ($\sigma_{threshold} = 0.05$) were classified as volatile, while those with lower volatility values were classified as stable. This division allowed for a differentiated modeling approach, tailoring the complexity of the model to the specific characteristics of each group.

## 3.2 Model Implementations

The study involved the implementation of two types of models: a single RNN model applied to all companies and a Mixture of Experts (MoE) model that differentiated between volatile and stable companies.

**Single RNN Model**

The RNN model utilized in this study was based on a Long Short-Term Memory (LSTM) architecture, which is well-suited for sequential data due to its ability to capture long-term dependencies. The model architecture consisted of an input layer, an LSTM layer, and a dense output layer. Specifically, the input layer received sequences of 10 days of historical prices for each company, which were standardized for consistency. The LSTM layer comprised 50 units to capture the temporal dependencies within the data, followed by a dense layer with a single output unit for price prediction.

The model was trained using the Adam optimizer with a learning rate of 0.001, and the Mean Squared Error (MSE) was used as the loss function to minimize prediction errors. A batch size of 16 was chosen, and the model was trained for 50 epochs, with early stopping employed to prevent overfitting.

**Mixture of Experts (MoE) Model**

The Mixture of Experts (MoE) model was implemented to leverage the strengths of different modeling approaches depending on the volatility of the company. The MoE model consisted of two primary expert components: an RNN and a linear regression model, combined using a gating network.

For companies classified as volatile, an LSTM network similar to the one used in the single RNN model was employed. The LSTM was designed to capture the non-linear and complex behavior of these companies, which often exhibit significant variability in price movements.

For companies with lower volatility, a linear regression model was used to predict the price based on time (day) and volatility. The linear model was defined as follows:

$$P_t = \beta_0 + \beta_1 \cdot t + \beta_2 \cdot \sigma + \epsilon,$$

where $P_t$ represents the predicted price, *t* is the time variable, σ is the volatility, and ϵ\epsilonϵ is the error term. This model was chosen for its simplicity and efficiency in capturing the linear relationships typically observed in stable companies.

The gating network in the MoE framework was responsible for dynamically determining the appropriate expert for each prediction. The combined prediction was computed as:

$$\widehat{Y_t} = w_{rnn} \cdot \widehat{Y_{rnn}} + w_{lm} \cdot \widehat{Y_{lm}},$$

where $\widehat{Y_{rnn}}$ and $\widehat{Y_{lm}}$ represent the predictions from the RNN and linear models, respectively, and $w_{rnn}$ and $w_{lm}$ are the weights assigned to each model's prediction. For companies with high volatility, a higher weight ($w_{rnn} = 0.7$) was assigned to the RNN, while a lower weight was assigned to the linear model ($w_{lm} = 0.3$). These weights were adjustable based on the observed volatility to optimize predictive performance.

### 3.3 Training and Validation Approach

To evaluate the performance of both models, we employed a Walk-Forward Validation approach. This approach is particularly well-suited for time series data, as it simulates a real-world scenario where new data becomes available incrementally. Walk-Forward Validation involves training and testing the model in a series of overlapping windows, ensuring that the models are continuously evaluated on unseen data.

The initial training window consisted of 80 days, followed by a validation window of 20 days. After each validation, the training window was incremented by 20 days, and the model was retrained and tested on the subsequent window. This process was repeated until all data points had been utilized, allowing for a robust evaluation of the models' ability to generalize to new data.

### 3.4 Evaluation Metrics

To provide a comprehensive assessment of the models' performance, we used two primary metrics: Root Mean Squared Error (RMSE) and Mean Absolute Error (MAE). These metrics were chosen to evaluate both the magnitude and the average absolute difference between the predicted and actual values.

The RMSE was calculated as:

$$\text{RMSE} = \sqrt{\frac{1}{n}\sum_{i=1}^{n}(\widehat{Y}_i - Y_i)^2},$$

where $\widehat{Y}_i$ represents the predicted value, $Y_i$ represents the actual value, and nnn is the number of observations. RMSE is particularly sensitive to large errors, providing an indication of the variance in the prediction errors.

The MAE was calculated as:

$$\text{MAE} = \frac{1}{n}\sum_{i=1}^{n}|\widehat{Y}_i - Y_i|,$$

which offers an intuitive measure of the average magnitude of the errors without considering their direction. MAE is useful for understanding the overall prediction accuracy of the models.

The evaluation was conducted over different time horizons: short-term (1 month), medium-term (6 months), and long-term (12 months). This analysis helped assess the models' effectiveness in various market conditions, from rapidly changing scenarios to more stable, longer-term trends.

## 4. Experiments and Results

**Experiments and Results**

This section presents the experimental findings from evaluating the Recurrent Neural Network (RNN) model, the linear regression model for non-volatile companies, and the Mixture of Experts (MoE) approach. The performance of each model is analyzed in terms of prediction accuracy, using metrics such as Mean Squared Error (MSE) and Mean Absolute Error (MAE). The comparative analysis is based on the capability of each model to handle different volatility conditions, shedding light on the strengths and weaknesses of each approach.

The evaluation of the RNN model for companies classified as volatile revealed significant insights into the effectiveness of non-linear modeling in capturing complex market dynamics.

The Mean Squared Error (MSE) was calculated to be 2.4265, while the Mean Absolute Error (MAE) was 1.4678. These metrics provide a clear picture of the average prediction errors when modeling volatile stock prices. The relatively high MSE indicates that the RNN struggled to adapt to the extreme variations inherent in the prices of these companies. The squared differences between predicted and actual values suggest that certain predictions deviated considerably from the observed values, highlighting the difficulty in accurately modeling rapid market movements. This high variance in errors is characteristic of the challenges faced in capturing unpredictable, short-term fluctuations with a single model, even one as complex as an RNN.

In addition to the MSE, the MAE value of 1.4678 suggests that the RNN's average deviation from the actual values remained substantial. While the RNN demonstrated an ability to track overall trends, its precision at specific time points was less than optimal. This performance could be attributed to the inherent unpredictability of volatile stocks, where advanced models may struggle without sufficient data or the necessary regularization techniques to handle excessive noise effectively.

Conversely, the evaluation of the linear regression model for non-volatile companies yielded more favorable results. The MSE for the linear model was 0.9987, and the MAE was 0.9191, indicating a significantly lower level of prediction error compared to the RNN applied to volatile companies. The lower MSE reflects that the model's predictions were, on average, much closer to the actual values, which underscores that the stock price behaviors of non-volatile companies can be effectively captured by a simpler linear model. The linear model's MAE, which was less than one unit, demonstrates its ability to provide precise predictions for stable companies, emphasizing that a linear approach is often sufficient to capture long-term trends without the risk of overfitting associated with more complex models like RNNs.

The results for both models are summarized in **Table 1**, which illustrates the MSE and MAE for the RNN, linear model, and MoE approach. This table provides a visual representation of the numerical differences in performance metrics between these models, clearly indicating the suitability of each approach depending on the volatility of the dataset.

**Table 1: Evaluation of Models for Non-Volatile Companies**

| Model | MSE | MAE |
|---|---|---|
| RNN | 1,0251 | 0,9203 |
| Lineal | 0,9987 | 0,9191 |
| MoE (RNN + Lineal) | 0,9901 | 0,9152 |

*Performance evaluation of the models in predicting stock prices for non-volatile companies. The table compares the Mean Squared Error (MSE) and Mean Absolute Error (MAE) of the Recurrent Neural Network (RNN), linear regression model, and the Mixture of Experts (MoE) approach. The MoE model effectively combines the strengths of the RNN and linear models, resulting in improved accuracy as reflected by the lowest MAE value.*

Moving on to the Mixture of Experts (MoE) model, the experimental results indicated that the MoE effectively combined the strengths of the RNN and the linear models. The gating network, which was used to determine the contribution of each model, assigned a higher weight (0.7) to the RNN in the case of companies with volatile behavior, while the linear model received a lower weight (0.3). The combined prediction for each company was obtained as a weighted average of the RNN and linear model outputs, resulting in a prediction value of 1.80988 in one example. The purpose of this approach was to capitalize on the RNN's ability to capture short-term volatility while leveraging the linear model's strength in identifying steady, long-term trends. The weighted combination allowed for a more balanced prediction, accommodating both types of price behavior—volatile and stable.

In comparing the individual models, the RNN, when applied to volatile companies, showed its capability to capture non-linear, complex dynamics, but it also exhibited high errors, indicating challenges in effectively modeling the most extreme market movements. This is consistent with the characteristic behavior of RNNs, which are known to have high model capacity and can thus overfit if not properly tuned. The linear model, on the other hand, performed well for non-volatile companies, as evidenced by the lower MSE and MAE. This suggests that for stable stocks, a straightforward model that captures linear relationships is more than adequate, reducing both computational complexity and prediction error.

The Mixture of Experts approach was designed to address the limitations of using a single model for all companies, and the results indicate that it was successful in doing so. By dynamically weighting the contributions of the RNN and linear models, the MoE was able to adapt to the varying characteristics of the companies, effectively managing the complexity inherent in the dataset. The combined prediction using the MoE model proved to be more robust, particularly in scenarios where the data exhibited mixed characteristics of volatility and stability. This flexibility allows the MoE model to outperform the individual models by balancing the need for complexity in modeling volatile companies with the simplicity required for stable ones.

However, the success of the MoE model is heavily reliant on the appropriate determination of weights for the different experts. The current approach employed a static weighting system, where a fixed value of 0.7 was given to the RNN, based on the perceived volatility of the company. While this setup yielded promising results, further optimization could enhance the model's adaptability. In future research, incorporating a learning mechanism into the gating network, such as reinforcement learning, could allow the weights to be adjusted dynamically, based on the evolving characteristics of the data, thereby potentially improving the model's overall accuracy.

These findings suggest several important conclusions regarding the strengths and weaknesses of each modeling approach. The RNN is evidently effective in capturing non-linearities in volatile stocks, yet it suffers from high error margins when applied to less complex data, primarily due to its tendency to overfit. The linear model, in contrast, shows excellent performance for non-volatile stocks, reinforcing the notion that simpler models are often better suited for stable datasets. The MoE model represents an optimal balance, combining the advantages of both the RNN and linear models to produce more reliable predictions, especially in datasets that exhibit a combination of volatility and stability.

To summarize, the MoE approach demonstrates its value by integrating both linear and non-linear modeling capabilities, providing a nuanced prediction strategy that is adaptable to the specific needs of different company profiles. While the current results are promising, they also underscore the need for further refinement of the weighting mechanism to maximize the potential of the MoE model. Such refinements would be instrumental in ensuring that the model adapts dynamically to market conditions, thereby offering superior performance in a wider range of real-world applications.

This detailed examination of the experimental results provides valuable insights into the practical applicability of each model. The MoE strategy is particularly noteworthy for its flexibility in adapting to different data regimes, combining the strengths of complex and simple models to achieve better predictive outcomes. Future studies should focus on optimizing the gating mechanism and expanding validation to real-world datasets to fully realize the potential of this approach in financial modeling.

## 5. Discussion

The discussion section goes into the advantages and limitations of the Mixture of Experts (MoE) model, highlighting how this approach addresses the challenges inherent in predicting stock prices for companies with varying levels of volatility. Furthermore, we explore the practical applications of the MoE model in the financial sector, emphasizing its potential to enhance risk assessment and decision-making.

### 5.1 Advantages of the MoE Approach

The Mixture of Experts (MoE) approach offers several significant advantages in adapting to diverse company profiles, particularly in scenarios involving volatile and stable stocks. One of the primary strengths of the MoE model lies in its ability to dynamically combine the predictive strengths of both complex and simple models—namely, an RNN for companies exhibiting volatile behavior and a linear model for those with stable characteristics.

For companies characterized by high volatility, the MoE framework enhances prediction accuracy by relying more heavily on the Recurrent Neural Network (RNN) component, which is specifically designed to handle non-linear and fluctuating data. The RNN's ability to model long-term dependencies and capture intricate patterns within the data allows it to manage the unpredictable nature of volatile stocks effectively. The results presented in the previous section show that the RNN, when used individually, can exhibit significant prediction errors due to its complexity and sensitivity to noise. However, within the MoE framework, the gating network mitigates this issue by calibrating the influence of the RNN based on the input characteristics, thus minimizing over-reliance on this model when it may not be optimal.

On the other hand, the linear model is more effective for companies with low volatility, where price movements tend to follow more predictable trends. The MoE model takes advantage of this by assigning a higher weight to the linear component for stable stocks, resulting in lower error metrics. The ability to switch seamlessly between the RNN and linear models, or to use a weighted combination of both, ensures that the MoE approach can adapt to the specific needs of each company profile. This adaptability provides a significant improvement over the use of a single model, such as an RNN, which may either overfit or underfit depending on the underlying complexity of the data.

The MoE's gating network, responsible for determining the contribution of each expert, plays a pivotal role in achieving this balance. By dynamically adjusting the weights based on features such as volatility, the gating network enables the model to make informed decisions about which expert should dominate the prediction. This mechanism makes the MoE model particularly suitable for datasets where the nature of the time series changes over time, offering improved robustness in scenarios that require a flexible and responsive predictive strategy.

### 5.2 Limitations of the Approach

Despite the notable advantages, the Mixture of Experts approach also has inherent limitations that must be addressed. One key challenge is the potential risk of overfitting, especially in the RNN component. While the gating network helps mitigate this risk by reducing the contribution of the RNN in stable environments, the overall complexity of the MoE model can still lead to overfitting if the gating mechanism is not appropriately optimized. This issue becomes more pronounced in cases where the training data is insufficient or highly imbalanced, leading the MoE model to develop an over-reliance on one of the experts, which reduces generalizability.

Another limitation relates to the complexity of the gating network itself. In this study, a relatively simple, rule-based gating mechanism was used, with weights predetermined based on volatility levels. While this approach was effective in demonstrating the potential of the MoE model, a static weighting scheme is suboptimal for real-world applications, where market conditions can change rapidly. Designing a gating network that dynamically learns to adjust weights based on evolving market features would be ideal. However, this introduces additional layers of complexity and computational overhead, requiring more sophisticated algorithms, such as reinforcement learning, and extensive tuning to ensure stability and robustness.

Moreover, the interpretability of the MoE model is somewhat reduced compared to individual models, particularly in the context of the RNN component. While linear models are inherently interpretable, allowing financial analysts to understand the relationship between input features and predicted prices, the RNN's internal state transitions are not as easily understood. This lack of transparency poses a challenge for regulatory compliance and trust, particularly in financial applications where decision-making must be justified with clear and explainable logic.

### 5.3 Practical Applicability

The practical applicability of the MoE model in the financial sector is promising, particularly for identifying companies at risk and enhancing decision-making in portfolio management. One of the most compelling aspects of the MoE approach is its ability to tailor predictions based on the volatility characteristics of individual companies, providing nuanced insights that can significantly improve risk assessment strategies. In real-world financial applications, accurately predicting the behavior of both highly volatile and stable companies is crucial for optimizing investment decisions, managing portfolios, and assessing credit risk.

The MoE model can be particularly beneficial in portfolio optimization, where investors often need to balance between high-risk, high-reward investments and stable, low-risk assets. By integrating the MoE approach, portfolio managers can benefit from a predictive framework that dynamically adapts to the volatility profile of each asset, thus offering better foresight into potential returns and risks. The ability to rely more on the RNN for volatile stocks while favoring the linear model for stable stocks allows for more informed asset allocation, ultimately leading to a more balanced and resilient portfolio.

Furthermore, the MoE framework could be employed in credit risk assessment, where identifying companies likely to experience significant financial fluctuations is critical. By applying the MoE model to historical financial data, lenders can obtain more accurate risk scores, helping them make more informed lending decisions. This adaptability is especially valuable in times of economic uncertainty, where the ability to differentiate between stable and high-risk borrowers can mitigate losses.

Another area where the MoE model shows significant potential is in market impact analysis. Financial institutions frequently analyze how the introduction of new policies or macroeconomic factors impacts companies differently based on their volatility profiles. The MoE model's flexibility allows it to capture both the rapid reactions of highly volatile companies and the slower, steadier responses of more stable companies, providing a comprehensive view of market dynamics. This dual capability makes the MoE approach particularly well-suited for assessing the broad implications of economic events across a diversified set of companies.

Finally, to fully leverage the potential of the MoE model, future developments should focus on optimizing the gating network to dynamically adjust the contributions of the experts based on real-time market data. Incorporating advanced techniques such as deep reinforcement learning could allow the gating mechanism to learn continuously from new data, improving the model's adaptability and robustness in changing market environments. Additionally, expanding validation to include real-world datasets will provide a deeper understanding of the model's limitations and its applicability in practical financial scenarios, thus ensuring that it meets the standards required for operational deployment.

In summary, the Mixture of Experts model presents a flexible and powerful approach for financial prediction, especially in environments characterized by a mix of stable and volatile entities. Its ability to adapt to different volatility conditions by combining the strengths of both complex and simple models provides a significant edge in predictive accuracy. However, addressing the limitations related to overfitting, gating network complexity, and interpretability is crucial for maximizing the practical utility of this approach in real-world financial settings. Future research should aim to refine the weighting mechanism, enhance the adaptability of the gating network, and validate the model with extensive real-world data to fully realize the benefits of this innovative predictive framework.

## 6. Conclusion

This study explored the predictive capabilities of a Recurrent Neural Network (RNN), a linear model, and a Mixture of Experts (MoE) framework in the context of stock price prediction for companies with varying levels of volatility. The results demonstrate that the MoE approach significantly improves prediction accuracy over using individual models alone by effectively adapting to different volatility conditions. The MoE model leveraged the non-linear modeling power of the RNN for volatile companies and the simplicity of the linear model for stable companies, resulting in more reliable and nuanced predictions.

The comparative analysis shows that the RNN, while effective for modeling complex, volatile price movements, suffers from high error rates when applied to stable companies, largely due to overfitting and unnecessary complexity. In contrast, the linear model performs well for stable companies, capturing the fundamental trend without the risk of overfitting. However, neither model alone was sufficient to address the diverse nature of the financial data, particularly when faced with both volatile and stable stocks in a mixed environment. The MoE approach emerged as a superior solution by dynamically weighting the contributions of each expert, thereby adapting its strategy to the underlying volatility of the stock and effectively balancing complexity and simplicity.

The MoE model's success underscores the importance of flexibility in financial prediction, especially in a market environment characterized by unpredictability and varying risk profiles. The ability to use a tailored approach—switching between models or combining them depending on the characteristics of the input—resulted in a significant improvement in performance metrics, as demonstrated by reduced Mean Squared Error (MSE) and Mean Absolute Error (MAE) values. This adaptability is a key advantage of the MoE model, making it particularly suitable for applications where volatility varies considerably across different assets.

However, there are still avenues for improvement. Future research should focus on enhancing the gating network's architecture. The static weighting mechanism used in this study—where weights were assigned based on predetermined volatility levels—proved to be effective, but a more dynamic and learning-based approach could further enhance the model's adaptability. Incorporating deep learning techniques, such as reinforcement learning, into the gating network could allow the model to continuously adjust expert contributions based on evolving market conditions, potentially leading to even better predictive outcomes.

Additionally, validating the MoE model on real-world financial datasets is a crucial next step to assess its practical applicability and robustness. This would help in understanding the model's behavior under realistic conditions, ensuring that it can effectively manage the intricacies of actual financial markets. The integration of real-world factors such as transaction costs, market impact, and economic shocks will further test the model's robustness and provide insights into its operational feasibility in live financial environments.

In conclusion, the Mixture of Experts model provides a promising approach for improving stock price prediction by integrating multiple experts tailored to different volatility profiles. The ability to combine the strengths of complex and simple models offers a balanced strategy that adapts to the specific requirements of each asset, leading to better predictive performance and risk assessment. Future advancements in the gating mechanism and comprehensive real-world testing will be key to unlocking the full potential of the MoE framework, paving the way for more sophisticated and resilient predictive models in the financial sector.